\documentclass[12pt]{article}
\usepackage{amsfonts}
\usepackage{amsmath}
\usepackage{amssymb}
\usepackage[T2A]{fontenc}
\usepackage[cp866]{inputenc}
\usepackage[dvips]{graphicx}
\usepackage{epsfig}

\begin{document}
\def\be{\begin{equation}}\def\ee{\end{equation}}\def\l{\label}
\def\0{\setcounter{equation}{0}}\def\b{\beta}\def\S{\Sigma}\def\C{\cite}
\def\r{\ref}\def\ba{\begin{eqnarray}}\def\ea{\end{eqnarray}}
\def\n{\nonumber}\def\R{\rho}\def\X{\Xi}\def\x{\xi}\def\la{\lambda}
\def\d{\delta}\def\s{\sigma}\def\f{\frac}\def\D{\Delta}\def\pa{\partial}
\def\Th{\Theta}\def\O{\Omega}\def\o{\omega}\def\th{\theta}\def\ga{\gamma}
\def\Ga{\Gamma}\def\t{\times}\def\h{\hat}\def\rar{\rightarrow}
\def\vp{\varphi}\def\inf{\infty}\def\le{\left}\def\ri{\right}
\def\foot{\footnote}\def\ve{\epsilon}\def\N{\bar{n}(s)}\def\cS{{\cal S}}
\def\k{\kappa}\def\sq{\sqrt{s}}\def\bx{{\bf x}}\def\La{\Lambda}
\def\bb{{\bf b}}\def\bq{{\bf q}}\def\cp{{\cal P}}\def\tg{\tilde{g}}
\def\cf{{\cal F}}\def\bN{{\bf N}}\def\Re{{\rm Re}}\def\Im{{\rm Im}}
\def\K{{\cal K}}\def\ep{\epsilon}\def\cd{{\cal d}}\def\co{\hat{\cal
O}} \def\j{{\h j}}\def\e{{\h e}}\def\F{{\bar{F}}}\def\cn{{\cal N}}
\def\P{\Phi}\def\p{\phi}\def\cd{\cdot}\def\L{{\cal L}}\def\U{{\cal U}}
\def\Z{{\cal Z}}\def\ep{\epsilon}\def\a{\alpha}\def\ru{{\rm u}}
\def\vep{\epsilon}\def\ix{\mathbf{x}}\def\vs{\varsigma}\def\z{\zeta}
\def\vr{\varrho}\def\Sp{\mathrm{Sp}}\def\mA{\mathbf A}

\begin{center}
{\Large\bf QCD multi-jet processes and the multiplicity asymptotics}

\vskip 0.5cm

{\large J.Manjavidze\foot{\it Institute of Physics, Tbilisi, Georgia
$\&$ JINR, Dubna, Russia} $\&$ A.Sissakian\foot{\it JINR, Dubna,
Russia}}
\end{center}

\begin{abstract}
{\footnotesize We will offer the formal arguments that the very high
multiplicity (VHM) events are defined by the heavy QCD jets and, with
exponential over multiplicity accuracy, the number of jets must
decrease with the increasing multiplicity. This must rise the mean
transverse momentum of secondaries with multiplicity. The
effectiveness of the LLA in the VHM domain is also discussed.}
\end{abstract} \vskip 1cm

{\large\bf 1.} The interest of experimentalists in the very high
multiplicity (VHM) physics has grown for the last few years \C{vhmw}.
But in spite of all efforts there has not been any good quantitative
theory of such events until now. The reason lies in the special
kinematics of VHM events: the produced particles have approximately
the same small momenta and for this reason the usage of LLA becomes
problematic.

The purpose of the present paper is to give the formal arguments that
(i) the hadron processes become hard in VHM domain and (ii) this
tendency to the events hardness is conserved in the asymptotics over
multiplicity.

{\large\bf 2.} It is useful to introduce the generating function \be
T(z,s;n_{max})=\sum_n^{\infty}z^n \s_n(s;n_{max}),~ \s_n(s;n_{max})=0
~{\rm if}~n>n_{max}, \l{1}\ee where $n_{max}= \sqrt s/m$ is the
maximal multiplicity of hadrons of mass $m$ at given CM energy $\sqrt
s$. It appears because of the energy-momentum conservation law. We
will assume that the total energy $\sqrt s$ can be arbitrary large.
The lepton and photon production is neglected.

It seems natural to omit the dependence from $n_{max}$ if $n$ is far
from it: in the opposite case the dependence on the phase-space
boundaries would be sizeable for all $n$. We will assume that $z$ is
so small that \be n\ll n_{max}\l{2}\ee are important in (\r{1}).
Assuming (\r{2}), we can consider $T(z,s)$ as an {\it entire
function} of $z$ in the future.

One may turn (\r{1}) and define $\s_n(s)$ through $T(z,s)$ by the
inverse Mellin transform: $$ \s_n(s)=\f{1}{2\pi i}\oint
\f{dz}{z^{n+1}}T(z,s).$$ This integral can be calculated for large
values of $n$ by the stationary phase method. The corresponding
saddle point is defined by the equation: \be n=z\f{\pa}{\pa z}\ln
T(z,s).\l{4}\ee Then the asymptotic estimation exists: \be
\s_n(s)\propto e^{-n\ln z_c(n,s)},\l{5}\ee where $z_c(n,s)$ is the
solution of (\r{4}). The estimation (\r{5}) exists if $ z_c(n,s)> 1,$
i.e. if $n$ is larger than the mean multiplicity $\bar{n}(s)$; at the
same time $n$ can not be too large, see (\r{2}). The estimation
(\r{5}) signifies that the quantity \be \mu'(n,s) =-\f{1}{n}\ln
\s_n(s)/\s_{tot}(s)\l{7}\ee is defined with the $O(1/n)$ accuracy
only by the solution $z_c(n,s)$.

As follows from (\r{4}), with the increasing multiplicity $z_c(n,s)$
must tend to the singularities $z_s$ of $T(z,s)$: \be z_c(n,s)\to
z_s~. \l{8}\ee We will assume that \C{ly} $T(z,s)$ is regular in the
complex $z$ plane inside the unit circle (otherwise $\s_n$ would be
the increasing function of $n$ inside the domain (\r{2})). Keeping in
mind (\r{8}), the estimation (\r{5}) means the following supposition:

{\bf (I)}. {\it The asymptotics over multiplicity is governed with
the exponential accuracy by the leftmost singularity over $z$.}

It was shown \C{moi1} that the multiperipheral kinematics leads to
$T(z,s)$ which is regular for the arbitrary finite $z$. For this
reason the $soft$ channel of hadrons production leads, as follows
from (\r{4}), to $\mu'(n,s)\simeq \ln z_c(n,s)$ increasing with $n$.
On the other hand the pQCD jets generating function gives the
singularity at $z_s(s)-1\sim 1/\bar{n}_j(s)>0,$ where the mean
multiplicity in the jet is $\bar{n}_j(s)$, $\ln\bar
{n}_j(s)\propto\sqrt{ \ln s}$ \C{ukawa}. Therefore, for $hard$
processes, $$\mu'(n,s)= 1/\bar {n}_j(s) +O(1/n),$$ i.e. $\mu'(n,s)$
is the $n$ independent quantity which $decreases$ with the increasing
energy.

We conclude that the VHM processes are mostly the hard processes and
at first sight they can be described by the pQCD. Whether this is
right or not is the question of our interest.

{\large\bf 3.} We will consider particles production in the "deep
inelastic scattering" (DIS) kinematics. Let $\cf_{ab}(x,q^2;\o)$ be
the generating functional: $$ \cf_{ab}(x,q^2;\o)= \sum_\nu\int
d\O_\nu(k) \prod_{i=1}^\nu \o^{r_i}(k_i^2)
\le|a^{r_1r_2...r_\nu}_{ab}(k_1,k_2,...,k_\nu) \ri|^2,$$ where
$a^{r_1r_2...r_\nu}_{ab}$ is the production amplitude of $\nu$
partons ($r_i=(q,\bar{q},g)$) with momenta $(k_1,k_2,...,k_\nu)$ in
the process of scattering of the parton $a$ on the parton $b$;
$d\O_\nu(k)$ is the phase space element; $\o^{r}(k^2)$ is the "probe
function", i.e. the correlation functions $$ N^{r_1r_2...
r_\nu}_{ab}(k_1^2,k_2^2,...,k_\nu^2;x,q^2)=\le.\prod_{i=1}^\nu
\f{\d}{\d \o^{r_i} (k_i^2)}\ln\cf_{ab}(x,q^2;\o)\ri|_{\o=1}.$$ The
generating functional is normalized on the DIS structure function
${\cal D}_{ab}(x,q^2)$, $$ \cf_{ab}(x,q^2; \o=1)={\cal
D}_{ab}(x,q^2),$$ which is described in the leading logarithm
approximation (LLA) by the ladder diagrams. We will consider the
approximation when the cutting line passes only through the "steps"
of the ladder diagram. In this case ${\cal D}_{ab}(x,q^2)$ has a
meaning of the $probability$ to find the parton $a$ in the parton
$b$.

Considering the VHM region, one must assume that the mass $|k_i|$ of
the produced parton is large. For instance $k_i^2\gg \la^2$, where
$\la$ is the virtuality of the parton of the pre-confinement phase.
Then \be \ln k_i^2=\ln|q_{i+1}^2|\le(1+O\le(\f{\ln(1/x_i)}
{\ln|q_{i+1}^2|}\ri)\ri). \l{15}\ee The LLA means that \be \la^2\ll
|q_{i+1}^2|\ll|q_{i}^2|\ll|q^2|.\l{14}\ee Therefore, $|k_i|$ are also
strongly ordered. The approximation (\r{15}) means that the produced
partons longitudinal momenta are small: \be
\ln(1/x_i)<\ln|q_{i+1}^2|.\l{} \ee

It is useful to consider the Laplace image over $\ln(1/x)$: \be
\cf_{ab}(x,q^2;\o)=\int \f{dj}{2\pi i}\le(\f{1}{x}\ri)^j{\bf
F}_{ab}(j,q^2;\o).\l{18}\ee Then, taking into account the above
mentioned conditions, one may find the DGLAP evolution equation: \be
t\f{\pa}{\pa t} {\bf F}_{ab}(j,t;\o)=\sum_{c,r}\vp_{ac}^r(j)
\o^r(t){\bf F}_{cb}(j,t;\o),\l{17}\ee where $t=\ln(|q^2|/\La^2)$, $$
\vp_{ac}^r(j)\equiv\vp_{ac}(j)= \int_0^1 dx~x^{j-1}~P_{ac}^r(x)$$ and
$P_{ac}^r(x)$ are the regular kernels of the Bete-Solpiter equation
for pQCD \C{grib}. The equation (\r{17}) coincides at $\o^r=1$ with
the habitual equation for Laplace transform of the structure function
${\cal D}_{ab}(x,q^2)$. While the equation (\r{17}) was being
derived, only one additional assumption had been used for our problem
$\o^r=\o^r(k^2)$.

The dominance of the gluon contributions for the case $x\ll1$ will be
used and for this reason all parton indexes will be omitted. One may
find the solution of (\r{17}) in terms of the $\nu$-gluon correlation
functions $N^{(\nu)}$. Omitting the $t$ dependence in the
renormalized constant $\a_s$, let us write: $${\bf F}(j,t;\o)={\bf
D}(j,t)\exp\le\{\sum_\nu\f{1}{\nu!}\int \prod_{i=1}^\nu
dt_i(\o(t_i)-1)N^{(\nu)}(t_1, t_2,...,t_\nu; x,t)\ri\}.$$ In the VHM
domain, where $x\ll1$ is important, one must consider $(j-1)\ll1$.
Then $$N^{(1)}(t_1;j,t)=\vp(j)\sim\f{1}{j-1}\gg1.$$ The second
correlator $$ N^{(2)}(t_1,t_2;j,t)=O\le(\max\le\{\le(\f{t_1}{t}
\ri)^{\vp(j)}, \le(\f{t_2}{t}\ri)^{\vp(j)},\le(\f{t_1}{t_2}\ri)^{
\vp(j)}\ri\}\ri) $$ is negligible at $(j-1)\ll1$. Therefore, in the
LLA, $${\bf F}(j,t;\o)={\bf D}(j,t)\exp\le\{\vp(j)\int_{t_0}^t
dt_1(\o(t_1)-1)\ri\}.$$ Taking $\o(t)=const$, one may find that ${\bf
F}(j,t;\o)$ has the Poisson distribution with the "mean multiplicity"
$\sim \vp(j)t.$

If the time necessary to confine a parton into the hadron is
$\sim(1/\la)$, then the parton of the mass (virtuality) $|k|\gg\la$
must decay on the partons of lower mass since its life time is
$\sim(1/|k|)<<(1/\la)$. This is the reason of the pQCD jets
formation.

If the quantity $$\o(t,z),~\o(t,1)=1,~t=\ln k^2/\la^2,$$ is the
generating function of the pre-confinement partons multiplicity
distribution $$\o_n(t)=\le.\f{\pa^n}{\pa z^n}\o(t,z)\ri|_{z=0},$$
then, as follows from derivation of ${\bf F}(j,t;\o)$, the quantity
\be{\bf F}(j,t;\o)={\bf D}(j,t)\exp\le\{\f{1}{j-1}\int_{t_0}^t
dt(\o(t,z)-1)\ri\} \l{25}\ee will describe the pre-confinement
partons multiplicity distribution in the frame of LLA.

Inserting (\r{25}) into the integral (\r{18}), one can find that, if
we will use the designation: $$\bar{\o}(t,z)\equiv\int_{t_0}^tdt~\o(t
,z),$$ $j-1=\{\bar{\o}(t,z)/\ln(1/x)\}^{1/2}$ are essential in it.
So, to justify the LLA approximation, one should assume that the
"mobility" \be\{{\ln(1/x)}/\bar{\o}(t,z)\}\gg1\l{27}\ee decreases
with $z$ or, it is the same, with the multiplicity $n$. This is the
reason why the LLA for considered DIS kinematics has a restricted
range of validity in the VHM region.

Nevertheless, in the frame of LLA conditions, as follows from
(\r{25}), the generating functional $\cf_{ab}(x,t;z)$ has the
following estimation:$\ln\cf_{ab}(x,t;z) \propto\{\ln(1/x)
\bar{\o}(t,z)\}^{1/2}.$ Therefore, since the coupling is a constant,
$\ln\cf_{ab} (x,t;z=1)=\ln{\cal D}_{ab}(x,t)\propto(t\ln(1/x)
)^{1/2}.$ This is a well known result.

{\large\bf 4.} The higher correlation functions $N^{r_1r_2...
r_\nu}_{ab}(t_1,t_2,...,t_\nu; x,q^2)$ must be taken into account if
$z$ is large: the smallness of this correlation functions can be
compensated by the large value of $\o^r(t,z)$. This leads to
dominance of the multi-jet processes with jets of approximately equal
masses. Such kinematics is outside of the LLA. Let us consider it in
detail.

It is known \C{basseto} that $\o(t,z)$ is singular at \be z_s-1\sim
1/\bar{n}_j(t), \l{30}\ee where $\bar{n}_j(t)$ is the mean
multiplicity  in the jet: $\ln\bar{n}_j(t)\sim\sqrt{t}.$ As follows
from (\r{5}), the singularity (\r{30}) gives the estimation:
$\mu'=1/\bar{n}_j(t) +O(1/n),$ i.e. the character of singularity is
not important with $O(1/n)$ accuracy.

This means that, for instance, the multi-jet event with $k>1$ gives
the same contribution to $\mu'$. The difference may appear only in
the pre-exponential factor. But actually the difference will appear
since the singularity moves with $t$, $z_s=z_s(t)$. Indeed, if the
energy conservation law is taken into account and if the produced $k$
jets have the same energy, then $\bar{n}_j^{(k)}(t)<\bar{n}_j(t).$
Consequently, $z_s^{(k})(t)>z_s(t)$ and $\mu'^{(k)}_j> \mu'_j.$ This
means that the heavy jet productions mechanism would dominate in the
VHM region in accordance with the proposition {\bf (I)}.

{\large\bf 5.} The above derived result means the increase of the
mean transverse energy with the multiplicity $n$. The continuation of
this result into the asymptotics over multiplicity leads to the
deduction that only two heavy jets with masses $\sim {\sqrt s}/2$ are
produced in the deep asymptotics over $n$. However, the value of such
a process would be extremely small, $\propto (\a_s(s)/s)$.

Therefore, according to {\bf (I)} it is most probable to expect the
increase of the transverse energy with multiplicity, the experimental
confirmation of this result was given in \C{gutay} , but apparently
the limiting case of two-jet kinematics will never be exceeded.

\vskip 0.4cm {\large \bf Acknowledgement}\footnotesize

We would like to thank our colleagues in the Lab. of Theor. Phys. of
JINR and especially V.Kadyshevsky for great interest in the described
problem. The issue of the LLA for the VHM region was fruitfully
discussed with E.Levin and L.Lipatov many a time. The help of
N.Dokalenko in preparation of this paper was important. We are also
grateful to them for their help.


\end{document}